**Translating Social Media Crisis Narratives into Road Network Utilization Metrics: The Case of 2020 Oklahoma Ice Storm**

**H M Imran Kays**
Ph.D. Candidate
School of Civil Engineering & Environmental Science
University of Oklahoma
202 W. Boyd St., Norman, OK 73019-1024
Email: imran_kays@ce.mist.ac.bd

**Khondhaker Al Momin**
Ph.D. Student
School of Civil Engineering & Environmental Science
University of Oklahoma
202 W. Boyd St., Norman, OK 73019-1024
Email: momin@ou.edu

**Menziwokuhle Bandise Thwala**
Undergraduate Research Assistant
School of Civil Engineering & Environmental Science
University of Oklahoma
202 W. Boyd St., Norman, OK 73019-1024
Email: menziwokuhle.b.thwala-1@ou.edu

**K.K. "Muralee" Muraleetharan, Ph.D., P.E., G.E.**
Kimmell-Bernard Chair in Engineering, David Ross Boyd and Presidential Professor
School of Civil Engineering and Environmental Science
University of Oklahoma
202 W. Boyd St., Norman, OK 73019-1024
E-mail: muralee@ou.edu

**Arif Mohaimin Sadri, Ph.D.**
Assistant Professor
School of Civil Engineering & Environmental Science
University of Oklahoma
202 W. Boyd St., Norman, OK 73019-1024
E-mail: sadri@ou.edu
(Corresponding Author)

**ABSTRACT**

Risk communication in times of disasters is complex, involving rapid and diverse communication in social networks (i.e. public and/or private agencies; local residents) as well as limited mobilization capacity and operational constraints of physical infrastructure networks. Despite a growing literature on infrastructure interdependencies and co-dependent social-physical systems, an in-depth understanding of how risk communication in online social networks weighs into physical infrastructure networks during a major disaster remains limited, let alone in compounding risk events. This study analyzes large-scale datasets of crisis mobility and activity-related social interactions and concerns available through social media (Twitter) for communities that were impacted by an ice storm (Oct. 2020) in Oklahoma. Compounded by the COVID-19 pandemic, Oklahoma residents faced this historic ice storm (Oct. 26, 2020-Oct. 29, 2020) that caused devastating traffic impacts (among others) due to excessive ice accumulation. By using Twitter's recently released academic Application Programming Interface (API) that provides complete and unbiased data, geotagged tweets (~210K) were collected covering the entire state of Oklahoma and ice storm related tweets (~14.2K) were considered. First, the study uses natural language processing and text quantification techniques to translate crisis narratives (i.e. tweets). Next, geo-tagged tweets are mapped into co-located road networks using traditional GIS techniques. Finally, insights are generated using network science theories and quantified social narratives to interpret different elements of road networks (e.g. local roads, freeways, etc.) for the Oklahoma communities that were impacted by the ice storm event during the pandemic.

**Keywords:** Interdependency, Transportation, System, Crisis Mobility, Social Media, Twitter, Ice Storm

## INTRODUCTION AND MOTIVATION

The number of disasters around the globe has increased by a factor of five over the 50-year period between 1970-2019 [1]. Climate change has intensified these hazardous events, the majority being weather related events, that are occurring more frequently over space and time compared to historical records. This has led to the development of multi-hazard events defined by the United Nations Office for Disaster Risk Reduction (UNDRR) as events that occur simultaneously, cumulatively, or cascading over spatial and temporal scales [2].

In a study on the impact of disasters on physical infrastructure, Oh et. al. found that power, communication and road infrastructures were damaged or destroyed under the force of events such as the 2005 Hurricane Katrina [3]. The study also revealed that there exists interdependencies between these physical infrastructures, with transportation infrastructure systems being the most critical [3]. Road network infrastructure efficiencies are also impacted by natural disasters with a reduced serviceability [3, 4]. Other studies investigate how disasters stimulate dynamic secondary effects on dependent infrastructures such as health, hazard relief and response, and communication [5-9].

Traditionally road network system monitoring and assessment involved in-field methods of data collection and extraction. Survey Questionnaires, traffic counts, witness reports, etc. are among the basic methods for analyzing road network conditions [10-12]. For example, the Arizona Department of Transportation (DOT) traditionally used devices such as Dynamic Message Signs (DMS), Highway Advisory Radio (HAR), Road Condition Reporting Systems (RCRS), among others [13]. These approaches allowed manual entry of road or lane closures, vehicle restrictions, roadwork, road conditions, and multiple variables that would impact road network systems [13]. These methods have become obsolete in the investigation of road network conditions during hazardous events due to concerns about human safety [14].

In 2020, the COVID-19 pandemic posed challenges to the research community by limiting in-field research, minimizing movement, and altering the aforementioned research methodology approaches. However, it opened frontiers to explore social media (which was limited earlier) as a key element of research in multidisciplinary fields such as road network analysis. Social media platforms (SMPs) allow users to share ideas, thoughts, and information through virtual networks in a timely manner [15]. Users willingly share real-time information on SMP about topics such as road conditions. The COVID-19 outbreak  resulted in  additional increase in social media users (29.7%), who spend significant amount of time (1-2 hours per day) on SMPs [15, 16]. This makes social media a viable source of data extraction and analysis for the study of public risk perception. This makes online social media a viable source of data extraction and analysis for the study of public risk perception.

Compounded by the COVID-19 pandemic, Oklahoma residents faced a historic ice storm during  October of 2020 (10/26/2020-10/29/2020) [17]. This ice storm caused devastating impacts with more than 300,000 people without power and excessive ice accumulation causing trees to break and block roads among others resulting in more than $26 million in total damage [17, 18]. These effects reduced Oklahoma road network serviceability and increased traffic congestion as well as road accidents as a consequence of the dire state of the roads. With the rise of data mining software, Application Programming Interfaces (APIs) such as those offered by, Twitter, SMPs have become contenders in mass data extraction and analysis as opposed to traditional methods. The goal of this study is to investigate the potential of social media data as a viable alternative to

traditional sources for assessing road network utilization efficiencies during a natural disaster. The study also considers the 2020 Oklahoma ice storm as a major risk event to serve this purpose. The specific aims of this study are outlined below:

- *To investigate how online social networks (i.e. social media) depend on road networks, both spatially and temporally, based on public risk communication and risk perception before, during, and after a major hazard event.*
- *To observe the spatio-temporal variation of hazardous events and quantify crisis narratives from online social network relevant to road network infrastructure systems.*
- *To visualize and analyze the quantified crisis narratives and using them for the prioritization of road networks.*

## BACKGROUND AND RELATED WORK

Risk communication in times of disasters is complex, involving rapid and diverse communication in social networks (i.e. public and/or private agencies; local residents) as well as limited mobilization capacity and operational constraints of physical infrastructure networks. Despite a growing literature on infrastructure interdependencies and co-dependent social-physical systems, an in-depth understanding of how risk communication in online social networks weighs into physical infrastructure networks during a major disaster remains limited, let alone in compounding risk events.

The emergence of social media platforms such as Facebook, Twitter, Instagram, etc. has been met by a spike in social media users of an excess of 4.62 billion people as of 2022 global data portal by Kerpois [15]. This equates to 59% of the total global population generating large datasets of valuable information stored in text, photos, and videos. Now. these SMP are acting like a global connecting point where instant response, feedbacks are available; news broadcasting is faster than ever; e.g. twitter alone generates 143 K trending tweets per second [19]. Hence the other side of SMP application is the crisis communication during natural disasters. Users and agencies can take benefit of SMP during crisis mobility and risk communication, though there is a chance of misinformation to spread [20]. Irrespective to these limitations, SMP is providing an alternative way to capture public perception related to any event. As such, there is a growing tendency of research using SMP in a variety of analyzing natural disasters, including fire [21], flood [22], tsunami [23], earthquake [23, 24] , and hurricane [25, 26].

In a recent study, the authors found, that the most common methods of sentiment analysis use involve Lexicon based methods, while for machine learning it is Naive Bayes methods [27]. These methods are further divided into SentiWordnet and Term Frequency-Inverse Document Frequency (TF-IDF), and Support Vector Machines (SVMs), respectfully. Other studies have investigated the use of social media in research studies [28-33]. Fen et al. studied the interdependencies between user perception and the public transport system using Twitter to improve the Malaysian transport system [34]. This study applied sentiment analysis methodologies to understand users' satisfaction with the Malaysian transportation system modes. These methodologies are Lexicons (Afinn, Bing, and syuzhet) and Machine learning techniques (SVM) [35]. The study focused mainly on investigating the interdependencies between modes of transport within the transportation network and the user's perception as described by the analysis of sentiment values. The study found that Afinn-SVM had an accuracy (76.77%) and precision (76.38%) in the classification of user perceptions into two categories – service and punctuality [34]. Other studies investigate the user of Twitter data in transportation analysis [36-38]. Although showing the efficiency of sentiment analysis as a data analysis method, Fen et al leave a knowledge gap towards the perception in the transportation network infrastructure. This aspect is crucial in applying sentiment values for the purpose of improving transport system. The study further excludes consideration of Hazardous events which may greatly affect the data analysis of sentiment towards transport network systems from a spatio-temporal perspective.

In a study of the 2013 Moore, Oklahoma tornado by Ukkusuri et al. (2013), the knowledge gap on the influence of Hazardous events on SMPs user perceptions during crisis events was investigated. Ukkusuri et al. used sentiment analysis to reveal temporal and spatial variations in public perception during hazardous events. The purpose of this was to provide valuable information on crisis awareness, response and preparedness [39]. The study used rule-based classification techniques to identify content categories such as damage, injury report, fund-raising, and support and consolation. This provides an insight into the spatial effects of hazardous events with respect to user perception. The study further identified a trend in negative sentiment post event occurrence hence providing an insight into a potential shift in user perspective towards road network systems infrastructure risk perception over time. Similar observations are also found in other studies [40-43]. Ukkusuri et al. provide valuable information into spatial and temporal variation of user risk perception based on sentiment analysis in the case of hazard events. This was for the purpose of crisis awareness, response and preparedness and was solely based on user perception towards the 2013 Moore, Oklahoma tornado. This leaves a knowledge gap with regards to the transportation network infrastructure which informs the nature in which agencies respond to these hazardous events.

From the observed literature, it is apparent that sentiment analysis and word classification are two of the most important tools in interpreting crisis narratives from SMPs. The literature further shows very few attempts at investigating the interdependencies of social media networks and transportation infrastructure with respect to risk communication. This is of great concern when taking into perspective the influence of hazard events (i.e. tornado, hurricane, wildfire) on road network infrastructure and public risk perception on SMPs. This study presents a comprehensive approach in using natural language processing techniques to investigate the extent of interdependencies between social networks and the networks of roads in the neighborhood. In particular, this study analyzes large-scale datasets of crisis mobility and activity-related social interactions and concerns available through social media (Twitter) for communities that were impacted by an ice storm (Oct. 2020) in Oklahoma. Compounded by the COVID-19 pandemic, Oklahoma residents faced this historic ice storm (Oct. 26, 2020-Oct. 29, 2020) that caused devastating traffic impacts (among others) due to excessive ice accumulation. The results of this study will serve to identify if the risk communication in social media can be used to develop viable metrics to improve road network infrastructure systems.

The conceptual framework of this study is shown in Figure 1. The top layer shows how the public and agencies (i.e. Oklahoma DOT) can connect and interact in online social networks (i.e. Twitter). The middle layer shows the damage state of a community that includes several households and the surrounding physical infrastructure (i.e. roads) due to a natural hazard event. The bottom layer represents the expected outcomes i.e. the classification of neighboring roads based on functionality; red for not accessible and green for accessible roads. And this classification is based on the public discussions in online social networks about the functionality of the roads.

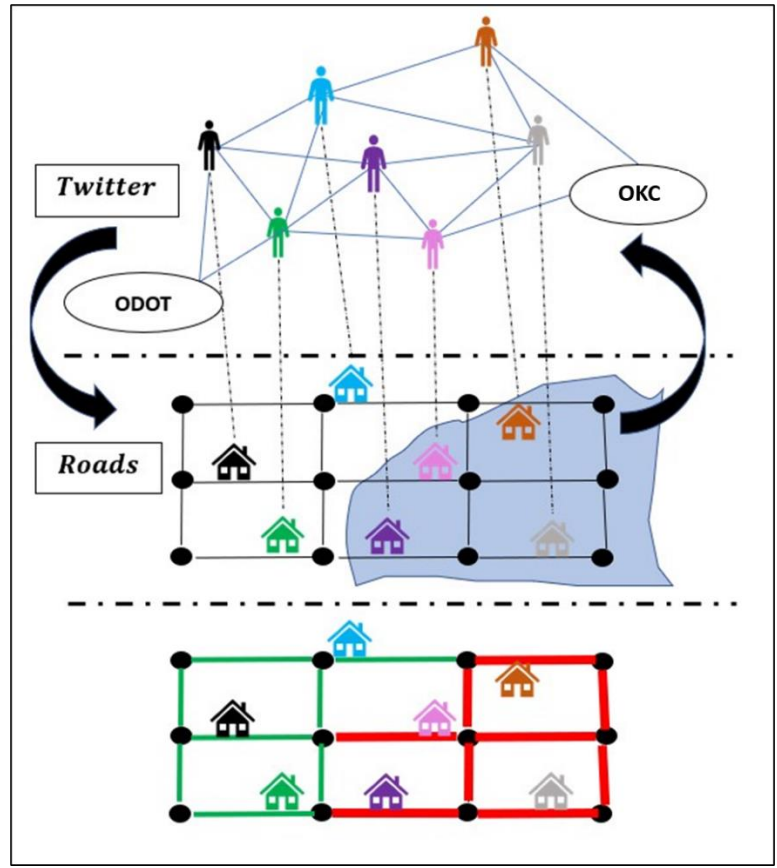

***Figure 1. Interdependency Interface between Roads and Online Social Networks***

## DATA DESCRIPTION

For this study, Twitter is identified as the SMP to capture the communication pattern before, during and after a natural hazard. The advantage is Twitter provides license for academic researchers and give access to public data for analysis. Moreover, Twitter academic API [44, 45] provides  full history of public conversations through full-archive search endpoint, within a boundary, within a timeframe [46, 47]. The time duration and location of data collection is selected focusing on historic October 2020 ice storm event in Oklahoma. Thirty-two days data from 10/19/2020 to 11/19/2020 is collected for whole state of Oklahoma. The "point radius" query option with a 25-mile radius circle is used to ensure that all location-based tweets are collected. The "point radius" query option generates a circle with a radius of 25 miles, as shown in Figure 2(a), and collect all tweets generated within the area. The API sets special constraints for geo-location based data collection i.e. the maximum length of the "point radius" cannot be greater than 25 miles [46]. To overcome this constraint, the entire state of Oklahoma is divided into 213 smaller squares (Figure 2(b)) with 25 miles on each side. Then, multiple "point radius" queries are utilized on these corner dots and generate circles of 25-mile radius. Though it generates substantial number of overlapped data (i.e. duplicate tweets), it covers the entire state of Oklahoma. Later the duplicate tweets are removed using machine learning technique.

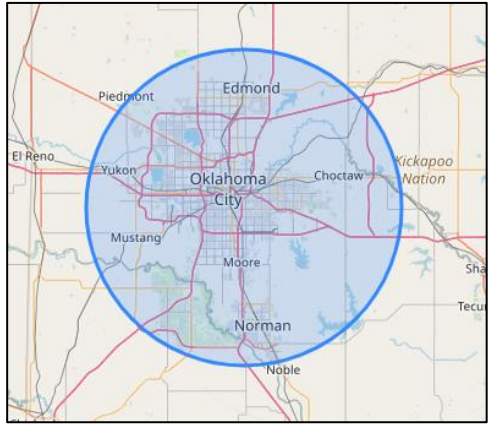

*Figure 2 (a) Spatial Context of "Point Radius" Query*

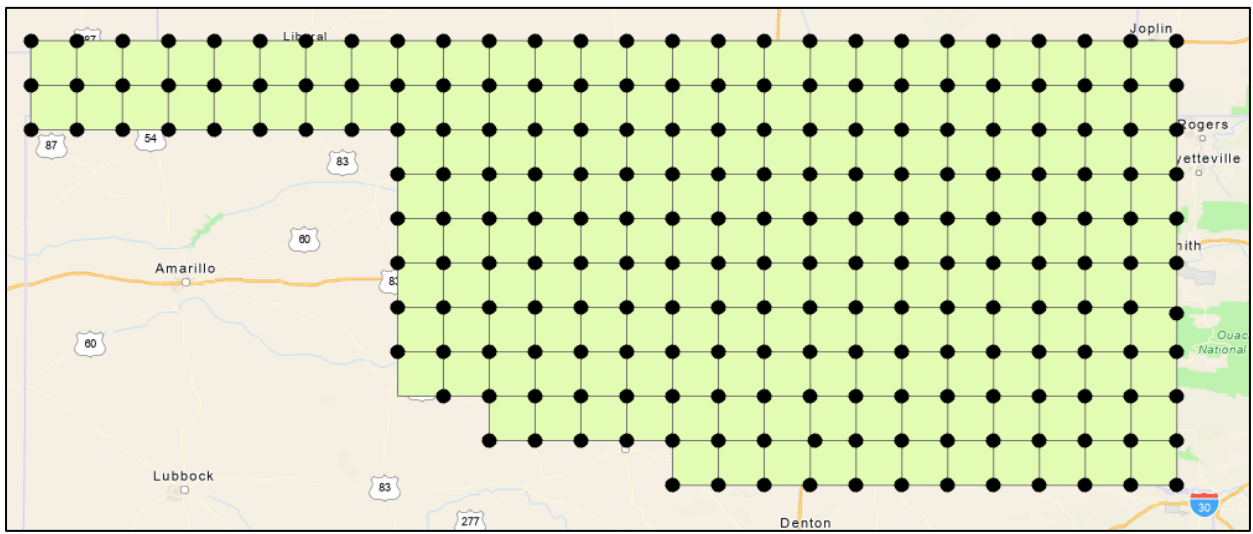

*Figure 2 (b) Center of Circles Location (25 miles) across the State of Oklahoma*

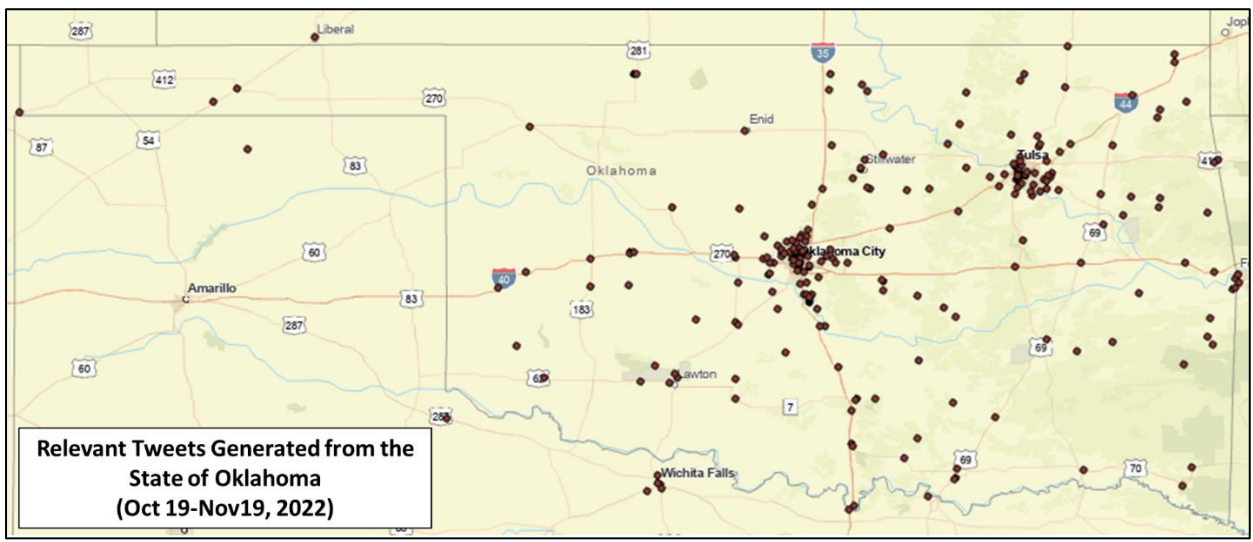

*Figure 3(a) Relevant Tweet Locations in the State of Oklahoma (total 14,220 tweets)*

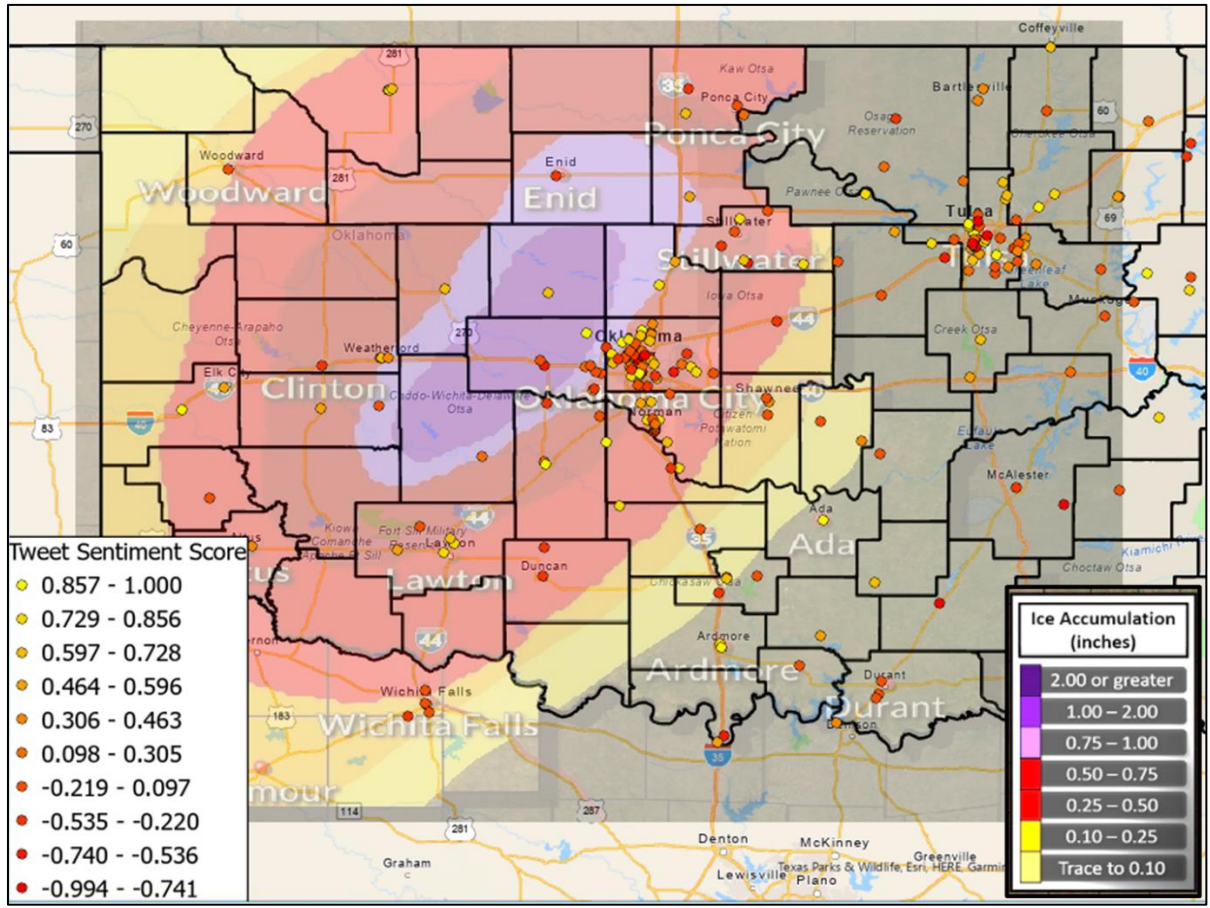

***Figure 3 (B) Locations of Tweets with Sentiment Score in Connection with Ice Accumulation Data [48].***

Overall, 420,000 tweets from approximately 15,500 unique users are collected. The '*Pandas'* library in Python is then used to check for repetitions and ensure that each tweet in the dataset are unique [49]. The study considers English tweets only: after filtering out non-English tweets, the raw data consists of approximately 210,000 unique geotagged tweets, all of which are in English. As the 25-mile point circle is also drawn on the corner points, it encompasses areas from neighboring states. Consequently, there were tweets from the neighboring states of Arkansas, Missouri, Kansas, New Mexico, Texas, and Colorado. The geotagged tweets were then screened by state, and in order to meet the objective of this study, only tweets from Oklahoma are considered. Elegant tweet preprocessing "*tweet-preprocessor*" Python package is used to remove noises: character codes, emojis, stop words, and html tags etc. Then, all the tweets are tokenized (broken down into smaller units: individual words or phrases). This study only considers tweets related to October 2020 ice storm event. The relevance of a tweet was determined by identifying tokens in the tweet; detailed steps can be found here [50]. Finally, 14,220 relevant tweets are identified relevant for the study. The locations of the relevant tweets are shows in Figure 3(a).

In addition to this, the tweet locations are compared with the weather map [48] by georeferencing the ice accumulation map in the same environment (Figure 3(b)). The weather map is provided by weather forecast office of Norman, Oklahoma where the ice accumulation are reported from 10/26/2020 to 10/29 2020 [48].

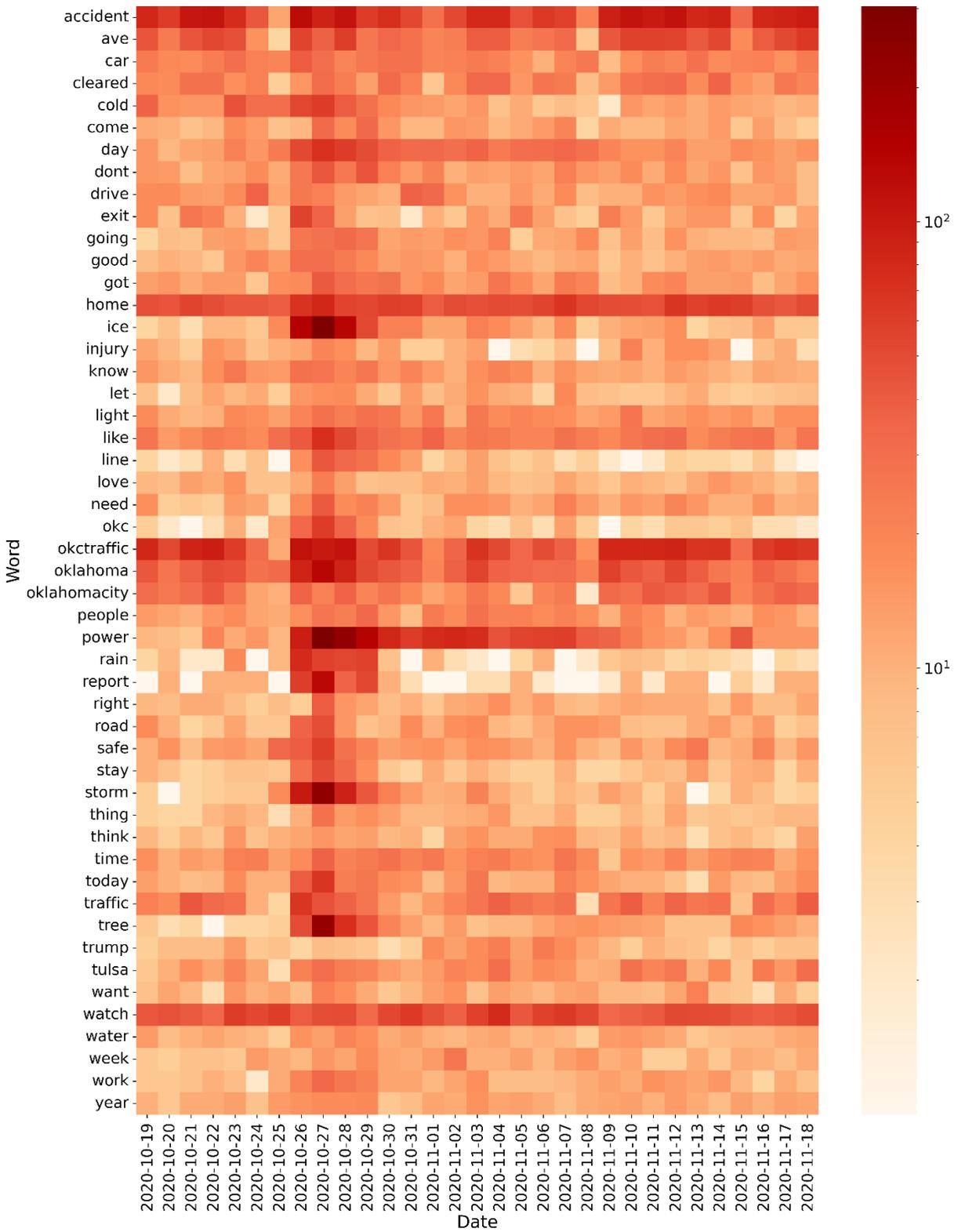

***Figure 4 Temporal Distribution of Top 50 Words in Oklahoma***

## Top 50 Frequent Words

The most frequently used words in a pool are sorted using frequent word analysis. The buzzwords in the texts tell the story. Heatmapping the most frequent words shows how they are stranded over time. It also shows how words rise and fall over time. Temporal heatmaps have been created for the top 50 most frequently occurring words (which is almost 26.45% of the total words) and illustrated in Figure 4. The words have been tallied in alphabetical order. These words can serve as a contextual "dictionary" as it incorporates all the aspects that can emerge in a discussion. The word frequency scale varies from the most frequent word "accident" (2314 mentions) to the 50th frequent word "injury" (293 mentions). The top five buzzwords are "accident" (2314 mentions), "okctraffic" (1923 mentions), "power" (1846 mentions), "home" (1717 mentions), and "watch" (1545 mentions). The heatmap's top frequent words reveal some obvious findings with the keywords: people frequently discussed the accident, property damage (i.e. house, tree), power outage, and traffic in Oklahoma.

## METHODS AND RESULTSRESULTS

### Translating Crisis Narratives using Sentiment Analysis

Sentiment analysis also referred to as opinion mining is defined as the task of finding the opinions of authors of the texts about specific entities [51]. Sentiment analysis is a form of natural language processing that involves emotion detection into three classification, negative, neutral and positive that exist within value ranges -1, 0 and +1 respectfully [51, 52]. Approaches to sentiment analysis can be further classified into machine learning approaches (MLA) and Lexicon based approaches (LA) [52].The MLA approaches apply famous ML algorithms and uses linguistic features such as negating words (e.g. "not") to calculate a sentiment score (S-score) related to a word group or a sentence [52]. Lexicon based approaches on the other hand involve the use of a collection of known and precompiled sentiment terms that have been given an S-score (i.e. the word 'death' may be allocated to an S-score of -0.7) [51, 52].

This study uses the *VaderSentiment* analysis tool that is based on the Lexicon approach of sentiment analysis [53]. Tweet data was cleaned in python environment to remove stop words and punctuations, returning the base words of twitter posts. These words, now defined as tokens, already have a sentiment score allocated to them from the *VaderSentiment* analysis library which were further computed to provide the complete sentiment of the tweet. A tweet word sentence would generally be defined as $X_w = (X_1, X_2, X_3..X_n)$, with each sentence elements (tokens) allocated to a value that would be used to compute the total sentence sentiment. Examples of tweets, their tokens and total sentence score are shown in the Table 1 below. In this study, estimated sentiment values for 14,220 tweets indicated the embedded emotions of twitter users towards the ice storm hazard. Geocoding these 14,220 tweets (Figure 3(b)) shows the tweet locations with different sentiment score in the map, where it is observed that, locations with higher accumulation of ice have more negative tweets.

**Table 1. Original Tweet, Tokens and Computed Sentiment Score**

| *Original Tweet* | *Tweet Tokens* | *Sentiment Score* |
|---|---|---|
| "This is such an Oklahoman sunset! The OU flag in front of the debris from the ice storm at sunset last night. #okwx #BoomerSooner" | ['oklahoman', 'sunset', 'flag', 'debris', 'ice', 'storm', 'night', 'okwx', 'boomersooner'] | 0 |
| "Good morning Oklahoma! @cityofokc starts the ice storm debris removal today."ðŸ˜," | ['good', 'morning', 'oklahoma', 'start', 'ice', 'storm', 'debris', 'removal', 'today'] | 0.6696 |
| "Please help feed hungry Oklahomans. I hear from people daily who struggle to eat because of unemployment... and recently, our devastating ice storm spoiled a lot of stored food." @reba Text FOOD to 501-501 to donate $10 | ['help', 'feed', 'hungry', 'oklahoman', 'hear', 'people', 'daily', 'struggle', 'eat', 'unemployment', 'recently', 'devastating', 'ice', 'storm', 'spoiled', 'lot', 'stored', 'food', 'text', 'food', 'donate'] | -0.6705 |
| With the pandemic and the recent catastrophic ice storm, this year has been extremely challenging for all Oklahomans. Join us in donating to United Way of Central Oklahoma. #supportlocal #community #leadership #downtownokc #okc #oklahoma #newmark | ['pandemic', 'recent', 'catastrophic', 'ice', 'storm', 'year', 'extremely', 'challenging', 'oklahoman', 'join', 'donating', 'united', 'way', 'central', 'oklahoma', 'supportlocal', 'community', 'leadership', 'downtownokc', 'okc', 'oklahoma', 'newmark'] | 0.4005 |
| OKC beginning massive debris pickup effort after last week's historic ice storm. Right now, 7+ days later, about 450 people in the city still have not had power restored https://t.co/e48ZPenkIB | ['okc', 'beginning', 'massive', 'debris', 'pickup', 'effort', 'week', 'historic', 'ice', 'storm', 'right', 'day', 'later', 'people', 'city', 'power', 'restored'] | -0.2584 |

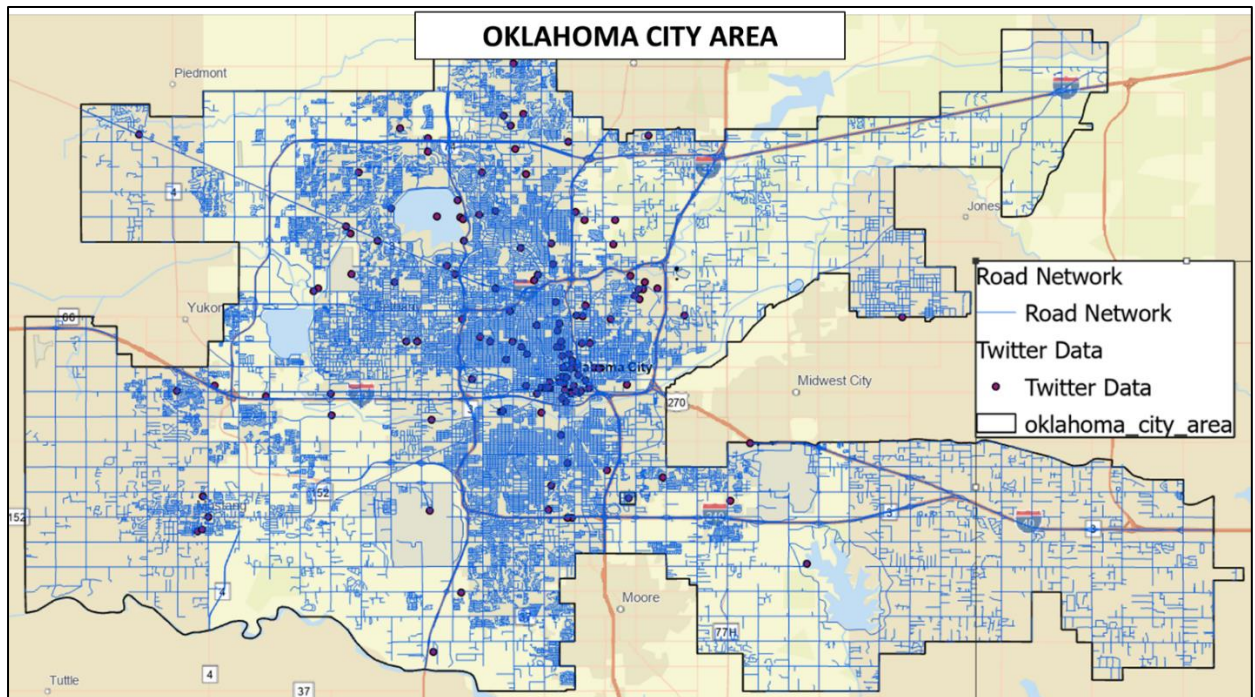

**Figure 5 Spatial Distribution of Tweets in OKC**

## Spatial Analysis

The collected 14,220 tweets are distributed all over Oklahoma, but it is observed that the density of tweets are higher around the city areas. To understand the communication pattern in twitter and relative condition in road network, smaller scale high resolution analysis will generate more reliable results. For this analysis, this study chosen Oklahoma City Area (OKC), where tweet density is higher than neighboring areas (tweet count 5,570). The locations are shown in figure 5. The tweet locations are clustered in one point due to the use of bounding box for data collection, yet many data points seem close to important business and residential district of OKC. For translating the sentiment score, roads of all functional classes i.e. the interstate highways, state highways, arterials, collectors, and local roads are considered for this analysis.

**Analysis of Influence Area**

For the analysis of twitter sentiment influence area in space, Thiessen polygon technique is used. This geographic technique is useful in identifying the zones under different spatial variables [54]. This method is extremely helpful in identifying space allocated for the influence of neighboring points [55]. This method is also known as weighted mean method, where the condition of defining each polygon is:

$$\bar{T} = \sum_{i=1}^{n} \frac{A_i}{A} T_i \tag{1}$$

Where, $T$ = twitter sentiment score, $A_i$ = Thiessen polygon area, $A$ = total area. Based on this principle, Thiessen polygon is generated inside the study area (OKC). The polygons are shown in Figure 6. Then the polygons are classified in 10 classes using Natural Breaks (Jenks) algorithm [56]. The color map in Figure 6 shows the location of influence areas with positive and negative sentiment values generated from tweet sentiment analysis.

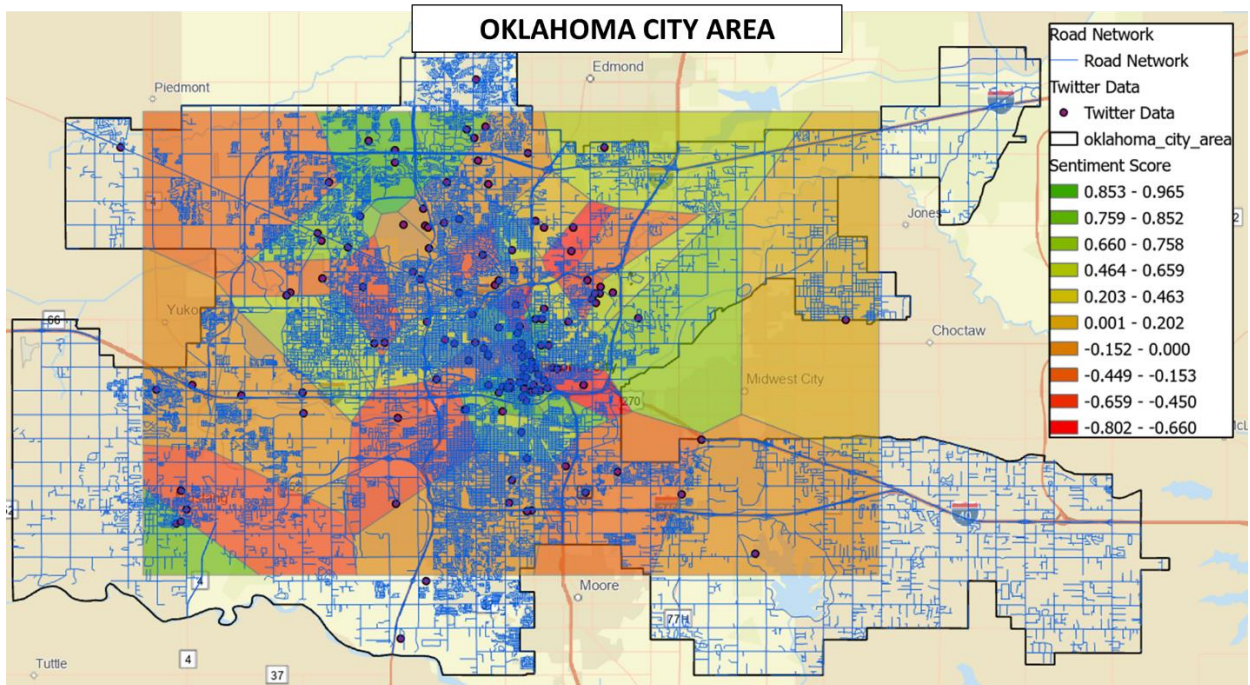

***Figure 6 Influence Areas (Thiessen Polygon) inside the Study Area (during October 26-29, 2020)***

**Mapping Sentiment Values on Road Networks**

In this analysis step, it is assumed that roads inside the influence area have such conditions that explains the reason of positive or negative sentiment of surrounding people. "Spatial join" tool in ArcGIS Pro V3.0 is used to identify the segments of road that falls inside the influence area [56]. Finally, the road segments are classified in 10 classes based on twitter sentiment score using Natural Breaks (Jenks) algorithm. The map in figure 7 (a) shows the road segments that are conveying the sentiment score from tweets. From Figure 3(b) it is found that OKC experienced 0.5 to 0.75 inches of ice accumulation during the ice storm. It is identified that areas close to critical zones (high ice accumulation area) tend to have road networks with more negative sentiment scores. This observation validates the hypothesis of this paper.

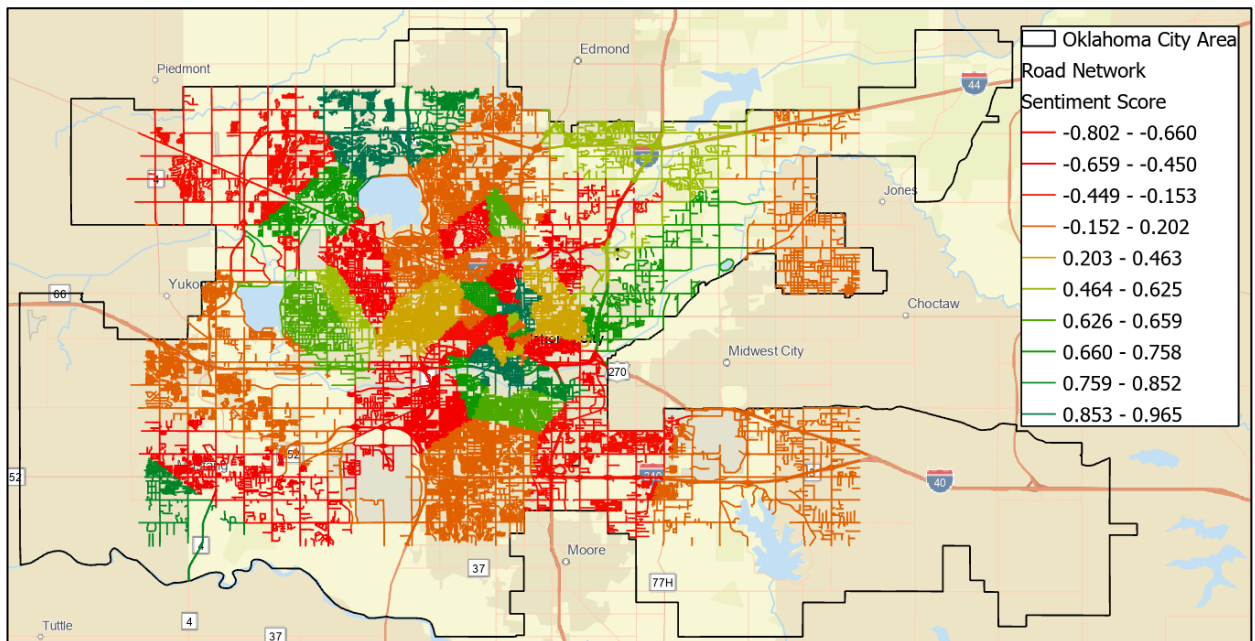

***Figure 7 Map Showing the Roads Reflecting the Sentiment Score from Tweets***

**APPLICATION**

This case study adopted the methodology described above, i.e. quantifying public sentiment and translate it to road network as a metric to identify the road condition during natural disasters. For validating the methodology, a small area of OKC is selected (Figure 8) which is surrounded by interstate highways (I-40, I-44, I-35). Other road classes considered within this area are arterials, collectors, and local roads; in general, roads of all functional classes are captured within this analysis area. In subsequent sections, this case study will investigate the influence of sentiment values estimated from tweet data during an ice storm event (October 2020 Oklahoma Ice Storm) over the topological credentials of the road network. Here, the topological credential of road network is defined as the identification of most critical and vulnerable components (e.g. road segment, intersections, etc.) by applying network topological metrics (Betweenness Centrality) [57].

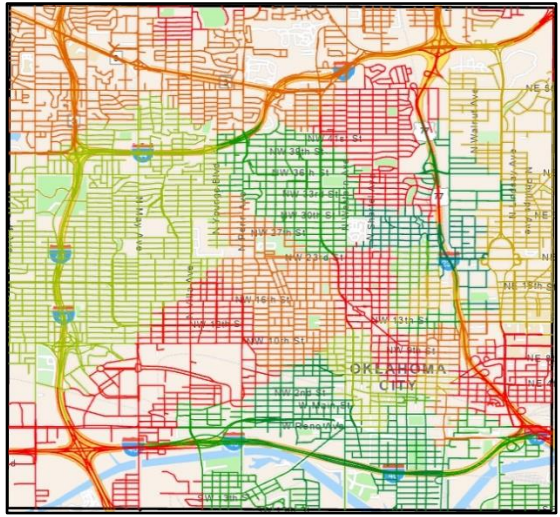

***Figure 8 Small part of OKC Road Network Selected for Case Study***

**Network Representation**

Topology of a network (*G*) consists of two components: nodes (*n*), i.e. the component itself and edges (*e*), i.e. the connection between the nodes. A sample network representation with node and edge are shown in Figure 9. Moreover, network can be directed of undirected. When the edges of a graph have the property to express the direction of information flow, it is defined as directed graph, e.g. one-way traffic; on the contrary if the edges represent both-way information flow, it is defined as undirected graph, e.g. both-way traffic. For this analysis, undirected graph is used due to simplicity of analysis. Moreover, graph can be weighted and unweighted, where the definition of weight varies depending upon the use of network, but it is assigned to nodes or edges to signify specific property inside it.

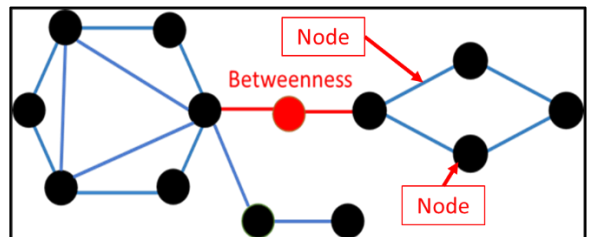

***Figure 9 Graph Representation with Nodes and Edges***

To represent the study area in graph format, the shape files are converted to nodes and edges using relational database in R where the geographic information is embedded [58]. In the study area, the nodes are defined as the road intersections and edges are defined as road segments between the nodes. Figures 9 and 10 illustrate the graph representation of the road network. After converting to graph, total number of nodes and edges are found as 4,924 and 8,452, respectively.

For this analysis, weighted graph is used where the weight is assigned to edges based on sentiment score estimated earlier. The weights are scaled between 0 and 1 is such a way that the more negative sentiment score gets highest value.

**Analysis of Topological Credential of the Network**

There are couple of ways to explain the topological credentials of a network. It can be either node level property or edge level property. Node level properties includes degree, closeness centrality, eigenvector centrality, betweenness centrality, PageRank centrality, etc. [59, 60]. On the other hand, edge level properties includes shortest path based metric, like, betweenness centrality [60]. For this case study, betweenness centrality (BC) for edges will be used. It is identified that edge betweenness centrality is one of the critical credentials for road network [57]. In general, betweenness centrality of edge is defined as the number of shortest paths in the network that pass through that node. Edge that most frequently sits on the shortest paths, gets higher BC value. BC value is a way to evaluate the amount of influence of an edge in the network based on shortest path (45). BC of edge in graph $G$ is expressed as:

$$BC\ (G) = \sum_{i,j} \frac{\gamma_{ij}\ (n)}{\gamma_{ij}} \tag{2}$$

Where, $\gamma_{ij}$ = number of shortest paths between node $i$ and $j$, $\gamma_{ij}\ (v)$ = number of shortest paths that pass-through node $v$ (not the $v$ as end point).

$$BC_w\ (G) = \sum_{j=1}^{n} A_{ij} w_{ij} \tag{3}$$

Where, $A_{ij}$ = the adjacency matrix of the network, $w_{ij}$ = is the weight matrix. In figure 9, the node and the edge marked in red sits on the most central location on the graph, hence have the highest BC value.

Based on Equation 2, the BC of the G is estimated for both weighted and unweighted conditions. The edges with unweighted BC value are shown in Figure 10. It is found that most of the central locations are on the highways surrounding the study area. This result is logical as the highways are providing connectivity for from-and-to city areas. Moreover, the highways are interconnected.

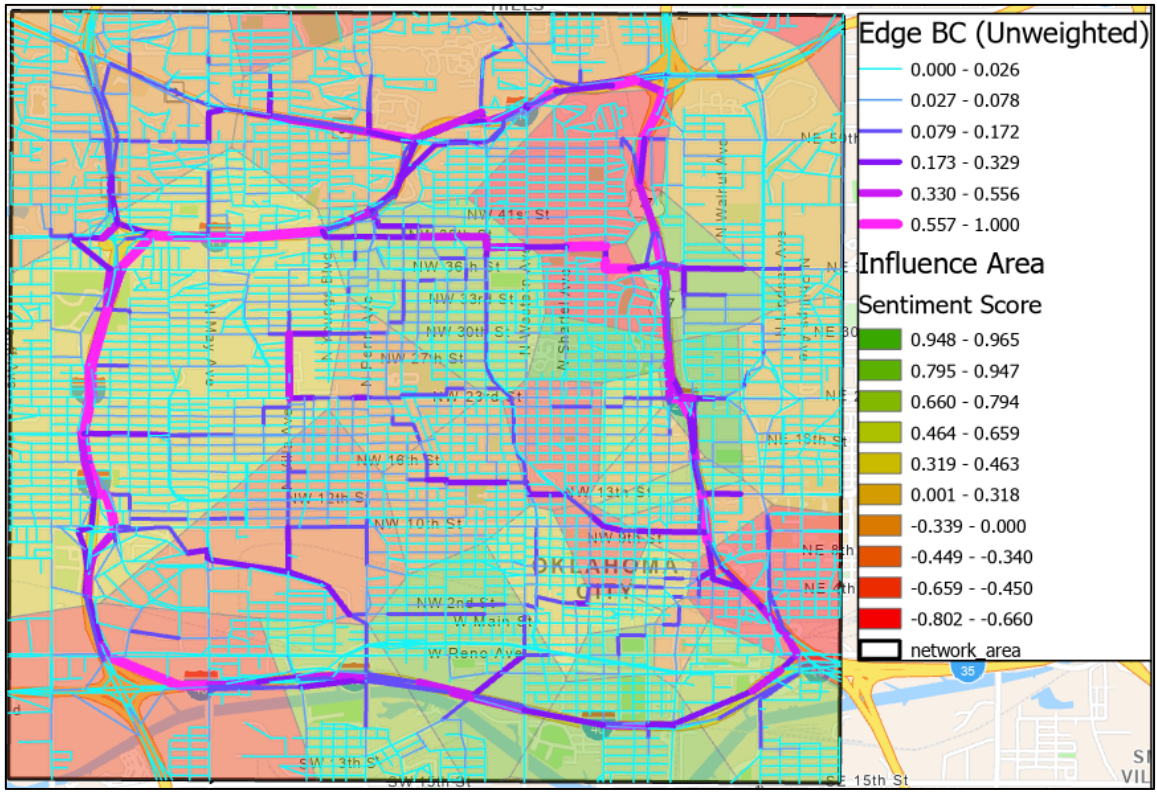

***Figure 10 Map Showing the Location of Edges in the Road Network (Unweighted BC)***

For the weighted BC analysis, Equation 3 is used to estimate the values. The results are shown in figure 11, where the road edges with weighted BC values are presented in the map. This analysis identifies some interesting facts. Firstly: the locations of the most central edges change after assigning weight to the analysis. While the highways have the most central edges in unweighted analysis, some of the local roads are having higher central positions in weighted analysis. This change is bought about by the incorporation of people's sentiment to the analysis. During the October 2020 ice storm, people posted tweets, where negative sentiments were mostly generated based on local condition. Hence the centrality (otherwise vulnerability) during the ice storm of the road network changed to more local roads from highways. Secondly: the northern side of the study region is the OKC uptown area, and the BC is changed from highways to local roads. The fact is, during ice storm people tweet based on localized context which shift the centrality, i.e. criticality of the roads from highways to local roads. Also, the south-west part of the study area has less population density, hence generated less tweets. The weighted BC indicates to the part of the study area where population density as well as the connectivity is high. In other words, these are the critical areas which requires priority attention during natural disaster and recovery period.

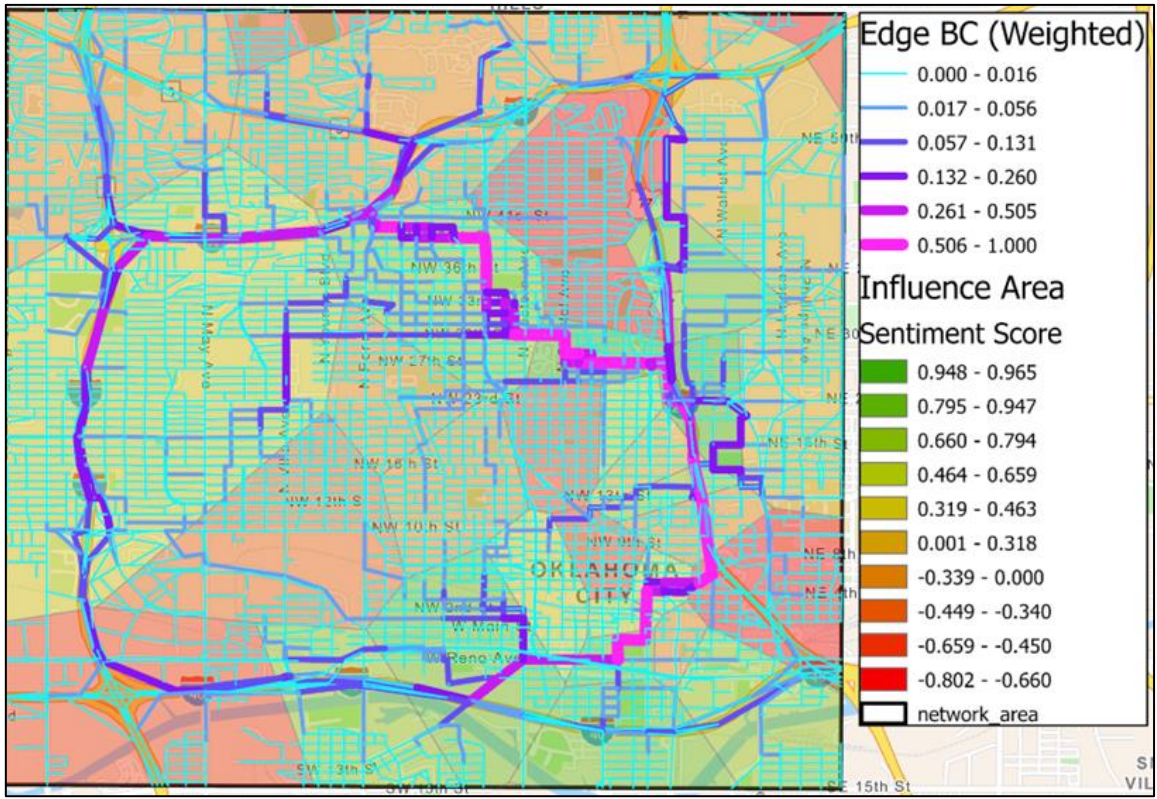

*Figure 11 Map Showing the Centrality of Edges in Road Network (Based on Weighted BC)*

## CONCLUSION AND FUTURE SCOPE OF RESEARCH

This research proposed a methodological framework to monitor the condition of a physical infrastructure based on public concerns and needs shared through social media platforms during a natural hazard event. The study contributes in the development of a systematic approach that translates social media narratives and utilizes these as metrics to assess the importance of the network components (i.e. topological credentials) of a physical infrastructure system. People tend to communicate frequently on social media during crisis especially when there is a risk of an approaching natural disaster. Moreover, with the tendency of sharing the sentiment with words, people share information about the surrounding infrastructures, their physical as well as operating conditions. Such information has the potential to unfold the system-wide performance of these infrastructures. Communicating these facts and figures with the local transportation and emergency management agencies would lead to better response in future disasters. The methodological framework developed in this paper can help by providing such information in disaster impacted communities. In addition to this, some key findings of this research are:

- *Crisis mobility and activity related interactions in online social media during natural disaster generate information (e.g. sentiment of user) based on the operating conditions of surrounding infrastructure (i.e. road network). Negative sentiments from the impacted residents are likely to generate from areas where the infrastructure components (e.g. roadways are blocked) are severely disrupted.*

- *Communication pattern (i.e. risk communication) in online social media can identify the vulnerable locations during a natural disaster. Such localized information regarding infrastructure disruptions due to disaster may not be captured by emergency management agencies. However, public risk communication patterns, if translated accordingly, can be quantified systematically to be used as a metric in identifying locations adversely impacted by a given disaster. For example, during Oklahoma 2020 ice storm event, negative sentiment of people (estimated based on tweets) in the impacted areas resembled locations with excessive ice accumulation.*
- *Priority of road segments also changes when sentiment values are treated as weights for the network components of roads. For example, while highways are more critical during normal operations, sentiment scores suggest that local roads require more attention during disasters.*

The proposed framework of this study creates a logical interface between social and physical systems based on information sharing. For example, communication pattern of Twitter (i.e. negative sentiment) during natural disaster identify the critical and vulnerable locations in road network, on the contrary locations with less vulnerability tends to generate positive tweets. However, it should be noted that social media platforms are not very popular in rural areas. For example, the tweets observed in the study area for the major Oklahoma ice storm (October 2020) were clustered mostly around the city areas. Hence the proposed data-driven methodology may not perform well enough for the rural areas. Also, there is a possibility that many tweets may not be geo-tagged, which may bias the outcome. Moreover, if the socio-demographic information of the Twitters users is not considered, it may create bias for certain age groups, gender, income group, among others. In addition, the spatial analysis is performed based on spatial data only, ecological fallacy is not tested.

Despite the aforementioned challenges and limitations, the proposed methodological framework has significant takeaways for future research. Some of these include but not limited to:

- Use of more advanced text classification may add more insights to vulnerability and criticality analysis.
- This methodology can be applied to a larger (i.e. state) or smaller (i.e. zip code) scale networks to examine the scaling effects.
- The methodology can be tested by integrating other online social media platform, like Facebook or Instagram, in addition to twitter.
- The research framework may further be extended from texts to images or videos shared by users on these social media platforms related to road network and other infrastructures.

**ACKNOWLEDGMENTS**

The material presented in this paper is based on work supported by the National Science Foundation under Grant No. OIA-1946093. The work of the second author was supported by NSF EPSCoR REU project. Any opinions, findings, and conclusions or recommendations expressed in this paper are those of the authors and do not necessarily reflect the views of the National Science Foundation.

## AUTHOR CONTRIBUTIONS

The authors confirm the contributions to the paper as follows: study conception and design: H Kays, A. M. Sadri; analysis and interpretation: H Kays, M Thwala, K Momin, A. M. Sadri; draft manuscript preparation: H Kays, M Thwala, K Momin, K Muraleetharan, A. M. Sadri. All authors reviewed the results and approved the final version of the manuscript.